\documentclass[10pt,twocolumn,letterpaper]{article}

\usepackage[pagenumbers]{cvpr} 
\definecolor{cvprblue}{rgb}{0.21,0.49,0.74}
\usepackage[pagebackref,breaklinks,colorlinks,allcolors=cvprblue]{hyperref}
\usepackage{xcolor} 
\usepackage{pifont} 
\usepackage{arydshln}
\usepackage{amsmath}     
\usepackage{amssymb}
\usepackage{tikz}   
\usepackage{multirow} 
\usepackage{tcolorbox}
\usepackage{graphicx}
\usepackage{multirow}
\usepackage{booktabs}
\usepackage{adjustbox}
\usepackage{colortbl}
\usepackage{bm}
\definecolor{cvprblue}{rgb}{0.21,0.49,0.74}
\definecolor{case1}{RGB}{255, 165, 0}
\definecolor{case2}{RGB}{162, 1, 1}
\definecolor{case3}{RGB}{0, 32, 96}
\definecolor{case4}{RGB}{25, 88, 105}
\definecolor{dog}{RGB}{195, 124, 11}
\definecolor{man}{RGB}{23, 86, 132}
\definecolor{grey_lc}{RGB}{154, 152, 153}
\definecolor{green_lc}{RGB}{93, 174, 86}
\definecolor{gray_lc2}{RGB}{89, 89, 89} 
\definecolor{gray_lc3}{RGB}{242, 242, 242}
\usepackage[pagebackref,breaklinks,colorlinks,allcolors=cvprblue]{hyperref}

\def\paperID{8611} 
\def\confName{CVPR}
\def\confYear{2025}

\title{Robust Audio-Visual Segmentation via Audio-Guided Visual \\ Convergent Alignment}
\author{Chen Liu$^{1,4}$, Peike Li$^{3}$, Liying Yang$^{5}$,  Dadong Wang$^{4}$, Lincheng Li$^{2}$,  Xin Yu$^{1}$\footnotemark[1]\\
$^{1}$ The University of Queensland, 
$^{2}$ NetEase Fuxi AI Lab,
$^{3}$ Matrix Verse AI, \\
$^{4}$ CSIRO Data61,
$^{5}$ Macau University of Science and Technology\\
{\tt\small\ yenanliu36@gmail.com, \tt\small\ xin.yu@uq.edu.au} 
}

\begin{document}
\maketitle
\renewcommand{\thefootnote}{\fnsymbol{footnote}}
\footnotetext[1]{Corresponding author.}
\begin{abstract}
Accurately localizing audible objects based on audio-visual cues is the core objective of audio-visual segmentation.
Most previous methods emphasize spatial or temporal multi-modal modeling, yet overlook challenges from ambiguous audio-visual correspondences—such as nearby visually similar but acoustically different objects and frequent shifts in objects' sounding status.
Consequently, they may struggle to reliably correlate audio and visual cues, leading to over- or under-segmentation.
To address these limitations, we propose a novel framework with two primary components: an audio-guided modality alignment (AMA) module and an uncertainty estimation (UE) module.
Instead of indiscriminately correlating audio-visual cues through a global attention mechanism, AMA performs audio-visual interactions within multiple groups and consolidates group features into compact representations based on their responsiveness to audio cues, effectively directing the model’s attention to audio-relevant areas. 
Leveraging contrastive learning, AMA further distinguishes sounding regions from silent areas by treating features with strong audio responses as positive samples and weaker responses as negatives.
Additionally, UE integrates spatial and temporal information to identify high-uncertainty regions caused by frequent changes in sound state, reducing prediction errors by lowering confidence in these areas.
Experimental results demonstrate that our approach achieves superior accuracy compared to existing state-of-the-art methods, particularly in challenging scenarios where traditional approaches struggle to maintain reliable segmentation.
\end{abstract}
\begin{figure}[t]
\begin{center}
\includegraphics[width=0.95\linewidth]{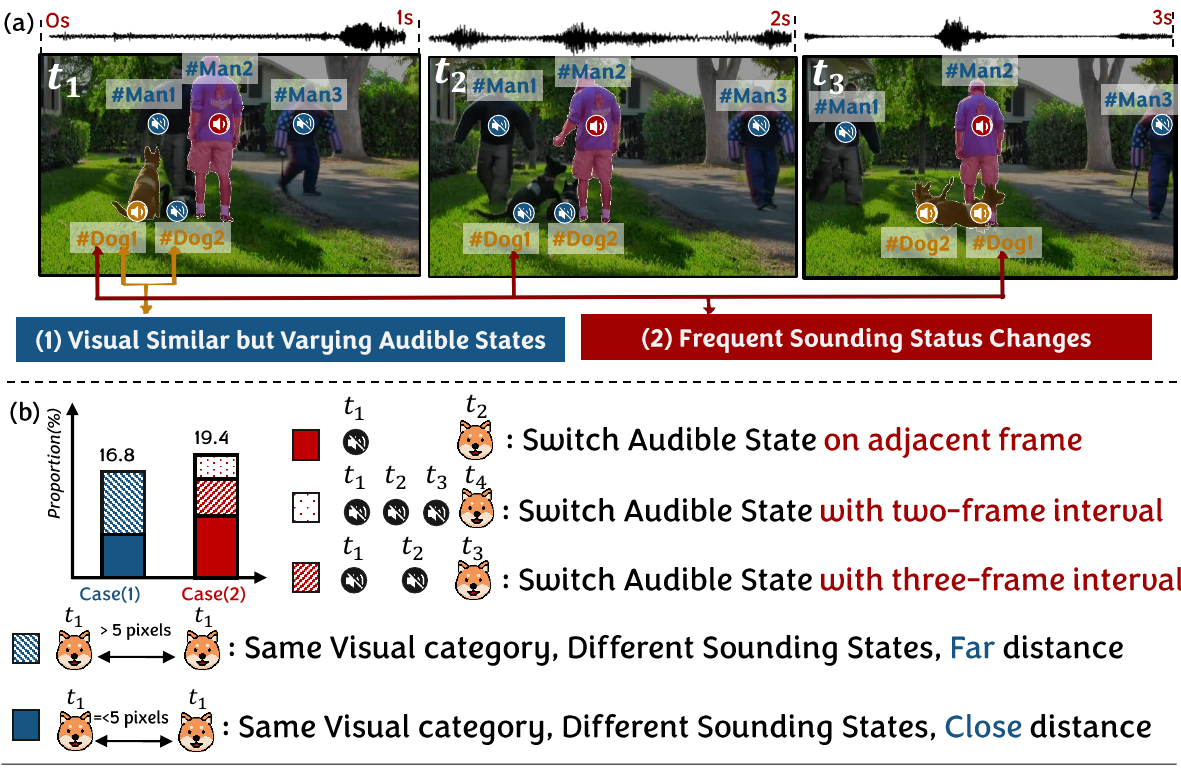}
\end{center}
\vspace{-1.5em}
\caption{\textbf{(a) Illustration of ambiguous spatio-temporal correspondences.} \textit{\textbf{\textcolor{case3}{Case (1):}}} At time $t_1$, \textbf{\textcolor{dog}{\#Dog1}} and \textbf{\textcolor{dog}{\#Dog2}} are positioned closely but have differing sounding states, challenging the model to identify the genuine sounding one.
\textit{\textbf{\textcolor{case2}{Case (2):}}} Over frames $t_1$, $t_2$, and $t_3$, \textbf{\textcolor{dog}{\#Dog1}} switches between sounding and silent states, posing challenges for models to capture the object’s sounding status variations over time reliably.
\textbf{(b) Distribution of Special Cases in AVSS Dataset.} We conduct a sample analysis utilizing a random 33.3\% subset of AVSS-V2 \cite{zhou2024audio}, which reveals substantial occurrences of cases (1) and (2), indicating the frequent presence of challenging frames.
}
\label{fig:teaser}
\end{figure}

\vspace{-1.0em}
\section{Introduction}
\label{sec:intro}
Given an audio signal, Audio-Visual Segmentation (AVS) aims to identify and segment audible objects within a visual scene \cite{zhou2022audio, zhou2024audio}.
Existing studies \cite{li2023catr, li2024qdformer, gao2024avsegformer} focus on modeling spatio-temporal audio-visual information to capture associations between audio and visual cues, such as dynamic sound properties and object motion.
Although these methods demonstrate strong performance, they often overlook challenges arising from ambiguous audio-visual spatio-temporal correspondences in real-world scenarios, potentially resulting in suboptimal segmentation, \emph{i.e.}, over- or under-segmentation.

As illustrated in Fig. \ref{fig:teaser}, we analyze and categorize the challenging cases in the AVS dataset, identifying two primary difficulties that arise from the audio-visual ambiguous correspondence.
\ding{182} \textit{\textbf{Incorrect Audio-Visual Associations Due to Visually Similar Objects with Different Sound States}} (\textbf{\textcolor{case3}{case 1}}).
In such cases, typical attention-based audio-visual alignment methods \cite{zhou2022audio, zhou2024audio, li2023catr} rely on global context to weight information. 
However, for visually similar yet acoustically distinct objects, these methods struggle to adjust attention distribution adaptively based on their actual sounding states, causing both silent and audible objects to be mistakenly aligned with audio cues.
To address this, some methods \cite{ma2024stepping, chen2024unraveling} employ ground-truth masks to filter out silent regions or employ ground-truth masks as references to construct positive and negative samples for contrastive learning. 
Although these approaches can temporarily exclude interference from silent areas, the model cannot rely solely on the GT mask to learn each object's unique sound response, resulting in attention still being mistakenly allocated to silent areas.
\ding{183} \textit{\textbf{Learning Instability Due to Frequent Sounding Status Changes}} (\textbf{\textcolor{case2}{case 2}}).
As shown in Fig. \ref{fig:teaser}, the sounding status of objects may frequently shift over short periods, complicating the learning of stable temporal audio-visual relationships.
When sound states change quickly, existing methods \cite{zhou2024audio, li2023catr, gao2024avsegformer, chen2024unraveling, ma2024stepping} tend to smooth over variations based on the preceding continuous sounding state, leading over-segmentation results.
In this paper, we propose a novel framework consisting of an audio-guided modality alignment (AMA) module aimed at addressing \ding{182} and an uncertainty estimation (UE) module targeting \ding{183}.

Considering that employing global attention-based methods may lead to dispersed attention, AMA first groups visual features based on their semantic density, thereby restricting audio-visual interaction to a group level. 
Then, the sound-guided semantic merging module performs audio-guided weighted merging of group features. 
Each feature is assigned a weight based on its audio responsiveness, with features corresponding to sounding objects being amplified and those corresponding to silent objects being suppressed. 
This process introduces an audio-driven feature competition that effectively differentiates sounding and silent objects, generating compact representations.
These representations are then reprojected onto the visual feature map.
After performing the above process across multiple layers, the sounding regions are highlighted while the silent regions gradually weaken.
To further strengthen the model's ability to differentiate between silent and sounding features, we treat compact representations with high audio responsiveness as positive samples and those with low responsiveness as negative samples. Contrastive learning is then employed to maximize the feature space distance between these positive and negative samples, ensuring better separation and discriminability \cite{chen2020simple, le2020contrastive, grill2020bootstrap, oord2018representation}.

Furthermore, to enhance the model's reliability in handling frequent sound state changes, we introduce an uncertainty estimation (UE) module. 
When an object's sound state shifts rapidly, the model may experience high uncertainty in mask predictions on transition frames. 
To address this, we first apply temporal modeling through an attention-based layer to capture temporal dependencies across frames. 
The resulting features are then fed into two separate heads in UE: a mask prediction head and an uncertainty estimation head. 
By incorporating uncertainty estimation into the mask prediction, the model adaptively reduces prediction confidence in regions with high uncertainty, effectively mitigating the impact of unreliable predictions during transitions.

In summary, our contributions are three-fold:
\begin{itemize}
    \item We introduce an audio-guided modality alignment module that adaptively identifies and highlights sounding regions through audio-driven feature competition and semantic grouping, improving the model's robustness against incorrect audio-visual associations.
    \item We develop an uncertainty estimation module to mitigate prediction errors caused by frequent changes in the sounding states of objects.
    \item Extensive experiments on benchmark datasets show that our framework outperforms previous state-of-the-art methods across all AVS datasets, with a 4.2\% gain in $\mathcal{J}\&\mathcal{F}_\beta$ on the AVS-Semantic dataset and an 11.5\% gain on the challenging VPO-MSMI dataset.
\end{itemize}

\section{Related Work}
\label{sec:formatting}

\subsection{Audio-Visual Alignment.}
Effective audio-visual alignment is crucial for achieving precise sound source segmentation.
Many studies have shown that attention-based modules \cite{zhou2024audio, li2023catr, gao2024avsegformer, li2024qdformer, ma2024stepping} and contrastive learning are powerful methods for modal alignment.
For instance, TPAVI \cite{zhou2024audio, zhou2022audio} utilizes an attention mechanism to align audio and visual modalities in both spatial and temporal dimensions.
To minimize interference from background noise and irrelevant regions, stepstone \cite{ma2024stepping} introduces ground-truth masks to filter out the silent visual regions, enhancing audio-visual alignment.
However, pre-filtering these silent regions prevents the model from learning to differentiate between silent and active objects, limiting its ability to fully explore multimodal integration.

Contrastive learning is another popular approach for audio-visual alignment, particularly in audio-visual localization \cite{mo2022localizing, sun2023learning, mo2023audio, mahmud2024t}.
Inspired by this, recent works \cite{chen2024cpm, chen2024unraveling} adapt contrastive learning to AVS to enhance audio-visual correlation. 
Traditional AVL methods \cite{tian2018audio, mo2022closer, rachavarapu2021localize, chen2021localizing, mo2022localizing, king2021audio, mo2023audio, masuyama2020self, hu2020discriminative, xia2022cross, qian2020multiple} treat audio-visual pairs from the same video as positive samples and pairs from different videos as negative.
However, this approach neglects the potential semantic alignment between features from different videos, mistakenly treating them as negative pairs and disregarding cross-video consistency.
To address this, \citet{chen2024unraveling} employ the ground truths to select positive samples, linking them with the corresponding audio signals, and treating background regions or mismatched audio-visual pairs as negative samples. 
While this method improves sample precision, it still overlooks semantic mismatches caused by visually similar but acoustically different objects in a frame.

In this work, we employ audio as the guidance to construct positive and negative samples, facilitating the model to be distinguishable for sounding and silent regions.

\vspace{-0.5em}
\subsection{Audio Visual Segmentation.}
Audio-visual segmentation aims to correctly partition the regions and label them with specific sound types \cite{zhou2022audio, zhou2024audio}.
Existing methods can be categorized by input granularity into two main types: frame-level input \cite{chen2024unraveling, chen2024cpm, liu2023audio, liu2024bavs} (\emph{i.e.}, a single frame paired with one-second of audio), and segment-level input \cite{li2024qdformer, zhou2022audio, zhou2024audio, mao2023multimodal, sun2024unveiling, liu2024annotation, huang2023discovering, gao2024avsegformer, hao2024improving, wang2024prompting, yan2024referred, yang2024cooperation, li2023catr}  (\emph{i.e.}, multiple frames paired with an audio segment of matching length).
For example, BAVS \cite{liu2024bavs} utilizes frame-level input to mitigate the model’s tendency to over-segment silent objects.
In contrast, methods emphasizing temporal information, such as CATR \cite{li2023catr}, claim that incorporating temporal dynamics better captures sound source variations over time, employing temporal audio-visual interactions through cross-attention mechanisms.
Moreover, with the emergence of query-based video segmentation models \cite{cheng2021mask2formervideo, ying2023ctvis}, several approaches have developed AVS frameworks built upon this architecture, incorporating audio information into the queries to facilitate audio-visual alignment \cite{ma2024stepping, li2024qdformer, sun2024unveiling}, thereby effectively modeling the temporal dynamics of the fused features.

While these approaches have improved AVS performance, there are still limitations. 
Frame-level methods fail to fully utilize multimodal information across time, limiting segmentation accuracy. 
Meanwhile, temporal modeling may favor generating smooth prediction results and overlook abrupt sound changes, resulting in suboptimal segmentation performance at sound transition points. 
In this work, we address these issues by estimating the prediction uncertainty caused by sound transitions and adjusting the mask prediction confidence based on the uncertainty map.




\begin{figure}[t]
\begin{center}
\includegraphics[width=0.95\linewidth]{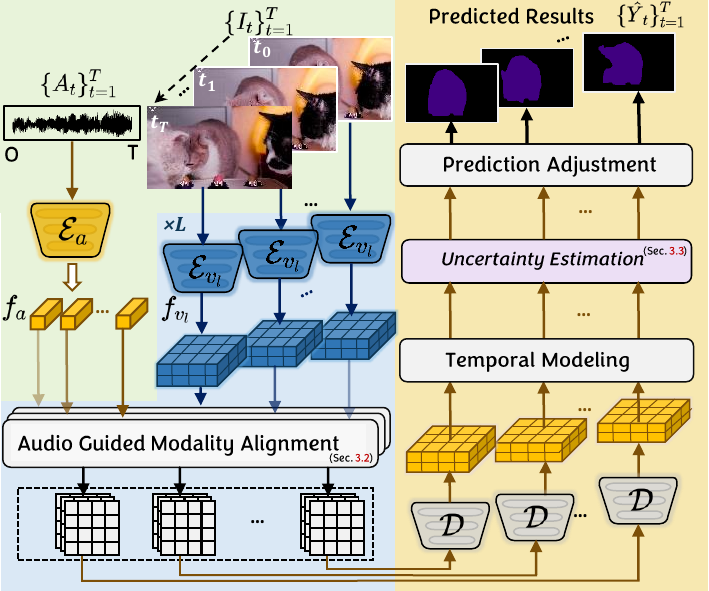}
\end{center}
\vspace{-1.5em}
\caption{\textbf{Method Overview.} Our framework takes video frames $\{I_t\}_{t=1}^T$ and audio signals $\{A_t\}_{t=1}^T$ as input to segment masks $\{\hat{Y}_t\}_{t=1}^T$ for audible objects. Visual and audio features extracted by the visual block $\mathcal{E}_{v_l}$ and audio encoder $\mathcal{E}_{a}$ are aligned through audio-guided modality alignment. The multi-scale features from each frame are then fed into the mask decoder to generate a fused feature map. Through temporal modeling, the feature maps are processed by the uncertainty estimation module to obtain the uncertainty map and mask confidence predictions. The final predicted results are generated by integrating the uncertainty map with mask confidence predictions. }
\label{fig:main_pipline}
\vspace{-1.5em}
\end{figure}

\section{Proposed Method}
\label{Sec:method}
Fig. \ref{fig:main_pipline} depicts a diagram of our proposed framework.
Before elaborating on our method, we first provide a preliminary in \S \ref{Sec:method_pre}. 
Then we introduce the key components of our method: \textit{\textbf{(i) Audio-Guided Modality Alignment}} (\emph{cf.} \S \ref{Sec:method_modalAlign}), which details the progress of facilitating the model in learning to distinguish between sounding and silent objects.
\textit{\textbf{(ii) Uncertainty Estimation}} (\emph{cf.} \S \ref{Sec:ST_Uncertain}), which measures the model's confidence changes over time regarding sounding state transitions.
The training objective is detailed in \S \ref{Sec:loss_func}.

\begin{figure*}[htp]
\begin{center}
\includegraphics[width=0.95\linewidth]{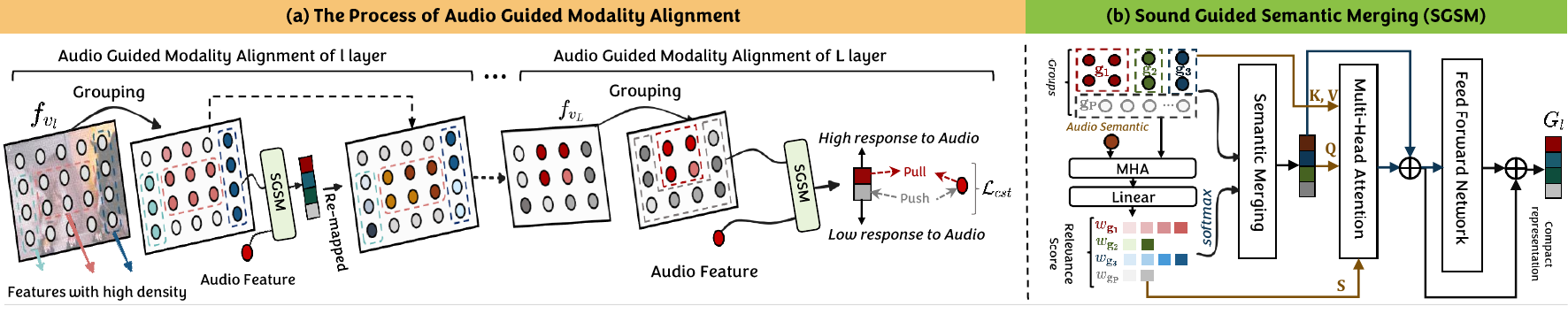}
\end{center}
\vspace{-2.0em}
\caption{\textbf{(a)} Image features are first grouped based on their semantic similarity. Audio and visual features interact at the group level, where features within each group are merged into compact representations guided by the audio signal. Through multiple layers of interaction, the sounding regions are progressively highlighted. The compact representations from the final layer are then used to perform contrastive learning with audio cues.
\textbf{(b)} Guided by audio features, the features within each group merge into compact semantic representations. These grouped semantics are then remapped onto the feature map to perform the next level of alignment.}
\label{fig:proalign}
\vspace{-1.5em}
\end{figure*}

\subsection{Notation and Preliminaries}
\label{Sec:method_pre}
\noindent\textbf{Task Setup.}
Formally, given a video consisting of $T$ one-second clips, where each second's data is represented as $\{I_t, A_t\}_{t=1}^T$, with $I_t$ denoting the last video frame of the $t$-th second and $A_t$ representing the corresponding one-second audio segment. 
A robust audio-visual segmentation model $f$ is designed to process the input pairs and generate pixel-wise localization maps $\{\hat{Y_t}\}_{t=1}^T$ for audible regions:
\begin{equation}
    \{\hat{Y_t}\}_{t=1}^T = f(\{I_t, A_t\}_{t=1}^T).
\end{equation}
The optimal solution $f^*$ is typically determined by minimizing a composite segmentation loss $\mathcal{L}_{SEG}$ across a dataset of $N$ videos $\{\mathcal{V}_n\}_{n=1}^N$:
\begin{equation}
   f^*=\underset{f}{\operatorname{argmin}}\frac{1}{NT}\sum_n\sum_t\mathcal{L}_{\text{seg}}(\hat{Y_t}, Y_t),
   \label{eq:seg_loss}
\end{equation}
where $\mathcal{L}_{\text{seg}}$ typically includes cross-entropy loss, Dice loss \cite{li2019dice}, and IoU loss \cite{yu2016unitbox}.

\noindent\textbf{Architecture Overview.}
Similar to other established audio-visual segmentation frameworks \cite{zhou2024audio, li2024qdformer, chen2024unraveling},  our architecture consists of a visual encoder with $L$ blocks, an audio encoder, and a mask decoder, defined as follows:
\begin{itemize}[label=\normalsize$\bullet$] 
    \item \textit{Visual Encoder} processes a single image $I_t \in \mathbb{R}^{H \times W \times 3}$ with resolution $H \times W$. Each block $\mathcal{E}_{v_l}$ extracts a feature map $f_{V_l} \in \mathbb{R}^{D_l \times H_l \times W_l}$, yielding a total of $L$ multi-scale feature maps, where $H_l$ and $W_l$ denote the spatial dimensions at level $l$, and $D_l$ is the number of channels.
    \item \textit{Audio encoder} $\mathcal{E}_a$ extracts an audio representation $f_a$ from the Mel spectrogram of the audio segment $A_t \in \mathbb{R}^{T \times F}$, where $T$ is the temporal duration and $F$ is the number of Mel filter banks. The output feature is $f_a = \mathcal{E}_a(A_t) \in \mathbb{R}^{D}$, with $D$ as the feature dimension.
    \item \textit{Mask decoder} $\mathcal{D}$ takes the multi-scale feature maps $\{f_l\}_{l=1}^L$ as input and produces a fused feature map $f_f \in \mathbb{R}^{D_f \times H_f \times W_f}$. Here, $H_f$ and $W_f$ denote the spatial dimension, and $D_f$ is the number of channels.
\end{itemize}

\subsection{Audio Guided Modality Alignment}
\label{Sec:method_modalAlign}
To focus the model's attention on sound-relevant regions and prevent it from spreading to irrelevant areas, we first group visual features by semantic similarity, and then merge these groups based on their responsiveness to sound, thereby achieving compact representations
Contrastive learning is subsequently applied to reinforce the distinction between sounding and silent objects by treating sound-responsive compact representations as positive samples and less responsive ones as negatives. 
Fig. \ref{fig:proalign} (a) illustrates the process intuitively.

\noindent\textbf{Visual Feature Grouping.}
We employ a k-nearest neighbor variant of Density Peaks Clustering (DPC-KNN) as \cite{du2016study, zeng2022not} for visual feature grouping due to its ability to effectively capture complex data distributions through local density.
Visual features are divided into $P$ groups, $\mathbf{G} = \{\mathbf{g_1}; \mathbf{g_2}; \dots; \mathbf{g_P}\}$, where $\mathbf{g_p} \in \mathbb{R}^{N_p \times D}$. 
Here, $N_p$ denotes the token amount in group $p$, which varies across groups.
Principally, DPC-KNN works as follows:

\noindent\ding{172} For each visual token $f_{V_l}^i$ (where $i = 1, 2, \dots, H_lW_l$), DPC-KNN begins by calculating the local density $\rho_i$ based on its $k$ nearest neighbors $\mathcal{N}_k(i)$:
\begin{equation}
    \rho_i = \sum_{j\in \mathcal{N}_k(i)} exp(-||f_{v_l}^i-f_{v_l}^j||).
\end{equation}
\noindent\ding{173} Each token $f^i_{V_l}$ is then assigned to a cluster by identifying its nearest neighbor $h_i$ with a higher density:
\begin{equation}
    h_i = \underset{j \in \mathcal{H}(i)}{\arg\min} \|f_{v_l}^i - f_{v_l}^j\|, \\
    c_i = \left\{
    \begin{array}{ll}
        i, & \text{if } \rho_i = \max(\rho) \\
        c_{h_i}, & \text{otherwise}
    \end{array},
    \right.
\end{equation}
where $\mathcal{H}(i)=\{j| \rho_i > \rho_j\}$ is the set of tokens with higher density than token $i$, and $c_i$ is the label assigned to token $i$.

With these assigned labels, we group the visual features, defining each group as $\mathbf{g_p}$ = $\{f_{v_l}^i|c_i=p\}$, $\forall p \in \{1, 2, \dots, P\}$.

\noindent\textbf{Sound Guided Semantic Merging.}
To prevent attention dispersion caused by global attention, we restrict the audio-visual attention calculation to local regions. 
We further merge visual features based on their responsiveness to audio signals, creating a compact representation that emphasizes features with high responsiveness to audio signals and more clearly reflects their association with audio cues, as depicted in Fig. \ref{fig:proalign} (b). 

To be specific, we first obtain the audio-visual interaction feature map $\hat{f_{v_l}}$ through a Multi-Head Cross-Attention (MCA) layer:
\begin{equation}
    \hat{f_{v_l}} = \text{MCA}(f_{v_l}, f_{a_l}),
\end{equation}
where $f_{a_l} \in \mathbb{R}^{D_l}$ is the audio feature projected through a linear layer to ensure dimensional consistency with $f_{v_l}$.
Next, we estimate the relevance score $S \in \mathbb{R}^{H_lW_l}$ of $\hat{f_{v_l}}$ via a multi-perception layer and calculate the weight vector $w_{\mathbf{g_p}} \in \mathbb{R}^{N_p}$ for each group by applying an exponential function.
This process is formulated as:
\begin{equation}
   w_{x} = \frac{\text{exp}(s_x)}{\sum_{j \in \mathbf{g_p}}\text{exp}(s_j)},\\
   g_p = \sum_{x \in \mathbf{g_p}} w_x \cdot x,
\end{equation}
where $\mathbf{g_p}$ is the set of visual tokens assigned to the group $p$, and $w_x$ is a scalar weight for $x  \in\mathbb{R}^{D_l}$ in group $p$. 
The compact representation of the input audio-visual pair is then defined as $G_l \in \mathbb{R}^{P \times D_l}$, where $P$ is the group number. This operation enables the model to focus on the most relevant tokens within each group, enhancing the distinction of regions with high audio response.

To highlight the sounding regions in the visual feature map, we enhance the compact representations using a query-based scheme and then project the updated representations back onto the visual feature map based on the token positions within each group. 
Specifically, the compact representations are updated by a stack of transformer decoders to converge high-level semantics:
\begin{equation}
    G_l \gets G_l + \texttt{softmax}(G_lf_{v_l}^T /\sqrt{D_l} + S)f_{v_l}.
\end{equation}
Here, the relevance score $S$ is added to the attention weights to ensure that visual tokens with high responsiveness to sound contribute more significantly to the output.
The feature map obtained by remapping the compact representations is then passed into the next block.
Specifically, we first retrieve the corresponding group feature from $G_l$ for each visual token $f^i_{v_l}$ based on its group index, and then add the retrieved feature to the visual token to enhance the representation of sounding regions.

\noindent\textbf{Contrastive-based Audio-visual Alignment.}
While the audio-guided feature merging yields compact visual representations that are more responsive to audio signals,  it does not fully separate silent objects from sounding ones in the feature space. 
To address this, we employ InfoNCE contrastive loss \cite{oord2018representation} to align audio and visual features for sounding objects while repelling those of silent objects.

Let $\hat{G} \in \mathbb{R}^{P \times D}$ represents the normalized compact features obtained in the final block, and $\hat{f}_a \in \mathbb{R}^{D}$ the normalized audio features. 
We compute similarity scores between audio and visual features via a dot product, followed by a sigmoid to derive alignment scores. 
These scores are then utilized to differentiate positive and negative samples based on their response to audio signals, as defined below:
\begin{equation} 
A = f_a \hat{G}^\top, I = \text{sigmoid}(A \cdot \sigma_p),
\end{equation}
where $A$ represents the similarity scores, and $\sigma_p$ is a scaling parameter that controls the sharpness of alignment.
The contrastive loss is then formulated as follows:
\begin{equation} 
\mathcal{L}_{\text{cst}} = -\log \frac{\sum_{i \in \mathcal{P}} \exp(A_i / \tau)}{\sum_{i \in \mathcal{P}} \exp(A_i / \tau) + \sum_{j \in \mathcal{N}} \exp(A_j / \tau)}.
\label{equ:cst}
\end{equation}
Here, $\mathcal{P}$ is the set of indices for positive samples, defined by $I_i > \epsilon_a$, $\mathcal{N}$ denotes the set of indices for negative samples, defined by $I_j < \epsilon_a$, and $\tau$ is the temperature parameter.

\subsection{Uncertainty Estimation.}
\label{Sec:ST_Uncertain}
When the sounding status of objects frequently changes in a video, the model may struggle to generate accurate and stable predictions during these transitions. 
To address this, we introduce uncertainty estimation to assess changes in the model's prediction confidence across all video frames. 
The final predictions are then refined by weighting uncertainty estimates alongside mask prediction probabilities.

\noindent\textbf{Generation of Uncertainty Map and Mask Probability Map.} After passing the multi-scale feature maps from all audio-visual pairs in a video through the mask decoder $\mathcal{D}$, we obtain fused feature maps. 
These fused maps are subsequently processed by a Multi-Head Attention layer to capture semantic dependencies across all time steps.
The resulting feature maps, denoted as $F_f = \{f_f\}_{t=1}^{T} \in \mathbb{R}^{T\times C\times H\times W}$, are input to the segmentation head, which generates the class logits $m \in \mathbb{R}^{T\times C\times H\times W}$ for each pixel.

To quantify the model’s prediction uncertainty, we adopt the Dirichlet distribution, as it provides granular, robust confidence information, enabling accurate uncertainty estimation and class competition \cite{minka2000estimating, sensoy2018evidential}. 
Specifically, $F_f$ is processed by the uncertainty head to obtain the uncertainty logits $\alpha \in \mathbb{R}^{T\times C\times H\times W}$. 
We then apply a softplus activation to the logits to ensure non-negativity. 
These values serve as the parameters for the Dirichlet distribution, which allows us to compute pixel-wise uncertainty as follows: 
\begin{equation} 
\delta = \frac{\alpha \left( \sum_{c=1}^{C} \alpha^c - \alpha \right)}{\left( \sum_{c=1}^{C} \alpha^c \right)^2 \left( \sum_{c=1}^{C} \alpha^c + 1 \right)}, \end{equation}
where $\delta \in \mathbb{R}^{T \times C \times H \times W}$ represents the prediction uncertainty across spatial locations. 
Higher values of $\delta$ indicate regions with greater uncertainty in predictions.

\noindent\textbf{Uncertainty-Weighted Prediction Adjustment.}
To ensure compatibility between uncertainty values and predicted probability distributions, we first normalize the spatial uncertainty by linearly scaling it to the [0, 1] range. 
The final predictions are then obtained via the following equation:
\begin{equation} \{\hat{Y}\}_{t=1}^T = {\sigma(m)}/({\delta + \epsilon}), \end{equation}
where $\sigma$ represents the softmax operation for semantic segmentation (or the sigmoid operation for binary segmentation), $m$ denotes the class logits from the segmentation head, and $\epsilon$ is set to $1 \times 10^{-6}$ to ensure numerical stability.

\subsection{Loss function}
\label{Sec:loss_func}
The final learning objective combines the typical segmentation loss $\mathcal{L}_{\text{seg}}$ (\emph{cf.} Eq. \ref{eq:seg_loss}) and the contrastive loss (\emph{cf.} Eq. \ref{equ:cst}):
\begin{equation}
    \mathcal{L} = \lambda_{\text{seg}} \mathcal{L}_{\text{seg}} + \lambda_{\text{cst}} \mathcal{L}_{\text{cst}},
\end{equation}
where $\lambda_{cst}$=0.1 and $\lambda_{seg}$=1 are the weight parameters.

\begin{table*}[htp]
\caption{Quantitative results on AVS-Objects and AVS-Semantic (resized to 224 $\times$ 224). The best results are highlighted in \textbf{bold}, while the second-best results are \underline{underlined}. \colorbox{gray_lc3}{\textcolor{gray_lc2}{Results}} are obtained using our complete design but with different backbones. See analysis in \S \ref{sec:quantitative_com}.}
\vspace{-0.5em}
\label{tab:avss}
\centering
\footnotesize
\setlength{\tabcolsep}{1.6mm}
\renewcommand\arraystretch{0.9}
\begin{tabular}{r|c|cccc|ccc|ccc}
\toprule
\multicolumn{1}{c|}{}                         & \multicolumn{1}{c|}{}                                  & \multicolumn{1}{c|}{}                                 & \multicolumn{3}{c|}{\textit{AVS-Objects-S4}}                                             & \multicolumn{3}{c|}{\textit{AVS-Objects-MS3}}                                            & \multicolumn{3}{c}{\textit{AVS-Semantic}}                                                 \\ \cmidrule{4-12} 
\multicolumn{1}{c|}{\multirow{-2}{*}{Methods}} & \multicolumn{1}{c|}{\multirow{-2}{*}{\shortstack{Visual \\ Backbone}}} & \multicolumn{1}{c|}{\multirow{-2}{*}{\shortstack{Audio \\ Backbone}}} & \multicolumn{1}{c}{$\mathcal{J}\&\mathcal{F}_m \uparrow$} & \multicolumn{1}{c}{$\mathcal{J} \uparrow$} & \multicolumn{1}{c|}{$\mathcal{F}_m \uparrow$}       & \multicolumn{1}{c}{$\mathcal{J}\&\mathcal{F}_m \uparrow$} & \multicolumn{1}{c}{$\mathcal{J} \uparrow$} & \multicolumn{1}{c|}{$\mathcal{F}_m \uparrow$}       & \multicolumn{1}{c}{$\mathcal{J}\&\mathcal{F}_m \uparrow$} & \multicolumn{1}{c}{$\mathcal{J} \uparrow$}  & \multicolumn{1}{c}{$\mathcal{F}_m \uparrow$}       \\ \midrule
TPAVI \cite{zhou2022audio}~\textcolor{gray}{\tiny{[ECCV22]}}                                         & PVT-V2-B5                                              & \multicolumn{1}{l|}{VGGish}                           & 83.3                     & 78.7                  & 87.9                         & 59.3                     & 54.0                  & 64.5                         & 32.5                     & 29.8                   & 35.2                         \\
AVSC \cite{liu2023audio}~\textcolor{gray}{\tiny{[ACM-MM23]}}                                         & Swin-Base                                              & \multicolumn{1}{l|}{VGGish}                           & 85.0                     & 81.3                  & 88.6                         & 62.6                     & 59.5                  & 65.8                         & \multicolumn{1}{c}{\_}   & \multicolumn{1}{c}{\_} & \multicolumn{1}{c}{\_}       \\
CATR \cite{li2023catr}~\textcolor{gray}{\tiny{[ACM-MM23]}}                                         & PVT-V2-B5                                              & \multicolumn{1}{l|}{VGGish}                           & 87.9                     & 84.4                  & 91.3                         & 68.6                     & 62.7                  & 74.5                         & 35.7                     & 32.8                   & 38.5                         \\
DiffusionAVS \cite{mao2023contrastive}~\textcolor{gray}{\tiny{[Arxiv23]}}                                  & PVT-V2-B5                                              & \multicolumn{1}{l|}{VGGish}                           & 85.7                     & 81.4                  & 90.0                         & 64.6                     & 58.2                  & 70.9                         & \multicolumn{1}{c}{\_}   & \multicolumn{1}{c}{\_} & \multicolumn{1}{c}{\_}       \\
ECMVAE \cite{mao2023multimodal}~\textcolor{gray}{\tiny{[ICCV23]}}                                       & PVT-V2-B5                                              & \multicolumn{1}{l|}{VGGish}                           & 85.9                     & 81.7                  & 90.1                         & 64.3                     & 57.8                  & 70.8                         & \multicolumn{1}{c}{\_}   & \multicolumn{1}{c}{\_} & \multicolumn{1}{c}{\_}       \\
AuTR \cite{liu2023audio_a}~\textcolor{gray}{\tiny{[Arxiv23]}}                                          & PVT-V2-B5                                              & \multicolumn{1}{l|}{VGGish}                           & 82.1                     & 77.6                  & 86.5                         & 72.0                     & 66.2                  & 77.7                         & \multicolumn{1}{c}{\_}   & \multicolumn{1}{c}{\_} & \multicolumn{1}{c}{\_}       \\
AQFormer \cite{huang2023discovering}~\textcolor{gray}{\tiny{[IJCAI23]}}                                      & PVT-V2-B5                                              & \multicolumn{1}{l|}{VGGish}                           & 85.5                     & 81.6                  & 89.4                         & 67.5                     & 62.2                  & 72.7                         & \multicolumn{1}{c}{\_}   & \multicolumn{1}{c}{\_} & \multicolumn{1}{c}{\_}       \\
AVSegFormer \cite{gao2024avsegformer}~\textcolor{gray}{\tiny{[AAAI24]}}                                  & PVT-V2-B5                                              & \multicolumn{1}{l|}{VGGish}                           & 86.8                     & 83.1                  & 90.5 & 67.2                     & 61.3                  & 73.0 & 40.1                     & 37.3                   & 42.8 \\
AVSBG \cite{hao2024improving}~\textcolor{gray}{\tiny{[AAAI24]}}                                        & PVT-V2-B5                                              & \multicolumn{1}{l|}{VGGish}                           & 86.1                     & 81.7                  & 90.4                         & 61.0                     & 55.1                  & 66.8                         & \multicolumn{1}{c}{\_}   & \multicolumn{1}{c}{\_} & \multicolumn{1}{c}{\_}       \\
GAVS \cite{wang2024prompting}~\textcolor{gray}{\tiny{[AAAI24]}}                                         & ViT-Base                                               & \multicolumn{1}{l|}{VGGish}                           & 85.1                     & 80.1                  & 90.0                         & 70.6                     & 63.7                  & 77.4                         & \multicolumn{1}{c}{\_}   & \multicolumn{1}{c}{\_} & \multicolumn{1}{c}{\_}       \\
BAVS \cite{liu2024bavs}~\textcolor{gray}{\tiny{[TMM24]}}                                         & Swin-Base                                              & \multicolumn{1}{l|}{Beats}                            & 86.2                     & 82.7                  & 89.8                         & 62.8                    & 59.6                  & 65.9                         & 35.6                     & 33.6                   & 37.5                         \\
COMBO \cite{yang2024cooperation}~\textcolor{gray}{\tiny{[CVPR24]}}                                         & PVT-V2-B5                                              & \multicolumn{1}{l|}{VGGish}                           & 88.3                     & 84.7                  & 91.9                         & 65.2                     & 59.2                  & 71.2                         & 44.1                     & 42.1                   & 46.1                         \\
QDFormer \cite{li2024qdformer}~\textcolor{gray}{\tiny{[CVPR24]}}                                      & Swin-Tiny                                              & \multicolumn{1}{l|}{VGGish}                           & 83.9                     & 79.5                  & 88.2                         & 64.0                     & 61.9                  & 66.1                         & \multicolumn{1}{c}{\_}   & 53.4                   & \multicolumn{1}{c}{\_}       \\
CAVP \cite{chen2024unraveling}~\textcolor{gray}{\tiny{[CVPR24]}}                                          & PVT-V2-B5                                              & \multicolumn{1}{l|}{VGGish}                           & \underline{90.5}                     & \underline{87.3}                 & \underline{93.6}                         & 72.7                     & \underline{67.3}                  & 78.1                         & 55.3                     & 48.6                   & 62.0                         \\
AVSStone  \cite{ma2024stepping}~\textcolor{gray}{\tiny{[ECCV24]}}                                     & Swin-Base                                              & \multicolumn{1}{l|}{VGGish}                           & 87.3                     & 83.2                  & 91.3                         & 72.5                     & \underline{67.3}                  & 77.6                         & \underline{61.5}                    & \underline{56.8}                   & \underline{66.2}                         \\
BiasAVS  \cite{sun2024unveiling}~\textcolor{gray}{\tiny{[ACM-MM24]}}                             & Swin-Base                                              & \multicolumn{1}{l|}{VGGish}                           & 88.2                     & 83.3                  & 93.0                         & \underline{74.0}                     & 67.2                  & \underline{80.8}                         & 47.2                     & 44.4                   & 49.9                         \\ \midrule
\rowcolor[HTML]{f2f2f2} 
\textcolor{gray_lc2}{\textbf{\textsc{RAVS (Ours)}}}                                          & \textcolor{gray_lc2}{PVT-V2-B5}                                                 & \multicolumn{1}{l|}{\textcolor{gray_lc2}{VGGish}}                           & \textcolor{gray_lc2}{91.3}                     & \textcolor{gray_lc2}{90.3}                  & \textcolor{gray_lc2}{92.2}                         & \textcolor{gray_lc2}{74.7}                     & \textcolor{gray_lc2}{69.7}                  & \textcolor{gray_lc2}{79.6}                        & \textcolor{gray_lc2}{63.9}                     & \textcolor{gray_lc2}{58.7}                   & \textcolor{gray_lc2}{69.0}                           \\
\rowcolor[HTML]{f2f2f2} 
\textcolor{gray_lc2}{\textbf{\textsc{RAVS (Ours)}}}                                          & \textcolor{gray_lc2}{MIT-B5}                                                 & \multicolumn{1}{l|}{\textcolor{gray_lc2}{VGGish}}                           & \textcolor{gray_lc2}{92.2}                     & \textcolor{gray_lc2}{91.6}                  & \textcolor{gray_lc2}{92.8}                         & \textcolor{gray_lc2}{74.2}                     & \textcolor{gray_lc2}{68.3}                  & \textcolor{gray_lc2}{80.1}                          & \multicolumn{1}{c}{\textcolor{gray_lc2}{\textbf{65.4}}} & \textcolor{gray_lc2}{\textbf{59.8}}                                 & \textcolor{gray_lc2}{\textbf{71.0}}                     \\
\rowcolor[HTML]{d9d9d9} 
{\textbf{\textsc{RAVS (Ours)}}}                                            & MIT-B5                                                & \multicolumn{1}{l|}{HTSAT}                                                    & \textbf{93.5}                                              & \textbf{93.1}                                  & \textbf{93.8}                                  & \textbf{76.4}                                     & \textbf{70.6}                                  & \textbf{82.1}                                & \textbf{65.7}                     & \textbf{60.8}                   & \textbf{70.6}                                       \\ \hline
\end{tabular}
\vspace{-1.5em}
\end{table*}


\section{Experiments} 
\label{sec:expriment}
\subsection{Experimental Setup}

\noindent\textbf{Datasets and Metrics.} We conduct experiments on: \emph{AVS-Object} \cite{zhou2022audio}, \emph{AVS-Semantic} \cite{zhou2024audio}, and \emph{VPO} \cite{chen2024unraveling}. 
\begin{itemize}
    \item \textbf{AVS-Objects} comprises 5,356 audio-visual pairs, each with five frames extracted from a video and a corresponding 5-second audio segment. Among these pairs, 4,932 contain a single audio event, while 424 involve multiple audio events. Binary masks are provided as ground truth to identify regions as either sounding or silent.
    \item \textbf{AVS-Semantic} extends the AVS-Object dataset, consisting of 12,356 audio-visual pairs. This extended dataset includes ten frames paired with a 10-second audio segment. It provides semantic annotations that identify both the sounding regions and the specific audio type.
    \item \textbf{VPO} is a synthetic dataset that combines single-frame images from COCO \cite{lin2014microsoft} with 3-second audio clips from VGGSound \cite{chen2020vggsound}. It includes three subsets: VPO-SS (12,202 single-source samples), VPO-MS (9,817 multi-source samples), and VPO-MSMI (12,855 samples), where weighted single-source data represent spatial information for multi-source sounds. 
\end{itemize}
Following \cite{li2024qdformer, chen2024unraveling, chen2024cpm}, we evaluate the performance of all methods using the mean Intersection over Union $\mathcal{J}$ and F-score ($\mathcal{F}_{\beta}$), where $\beta$ is set as 0.3.
Additionally, due to space constraints, we provide comparisons using the metrics from TPAVI \cite{zhou2024audio, zhou2022audio} in the \textit{supplementary material}.

\noindent{\textbf{Network Configuration.}} Our framework is end-to-end trainable, with all components parameterized by neural networks.
The spatial dimensions of the three feature maps used for sound-guided semantic merging are 1/4, 1/8, and 1/16 of the input image size.
The number of groups for each feature map is set to 14, 7, and 5. 
The threshold $\epsilon_a$  for selecting positive and negative samples is set to 0.5.

\begin{figure*}[htp]
\begin{center}
\includegraphics[width=0.9\linewidth]{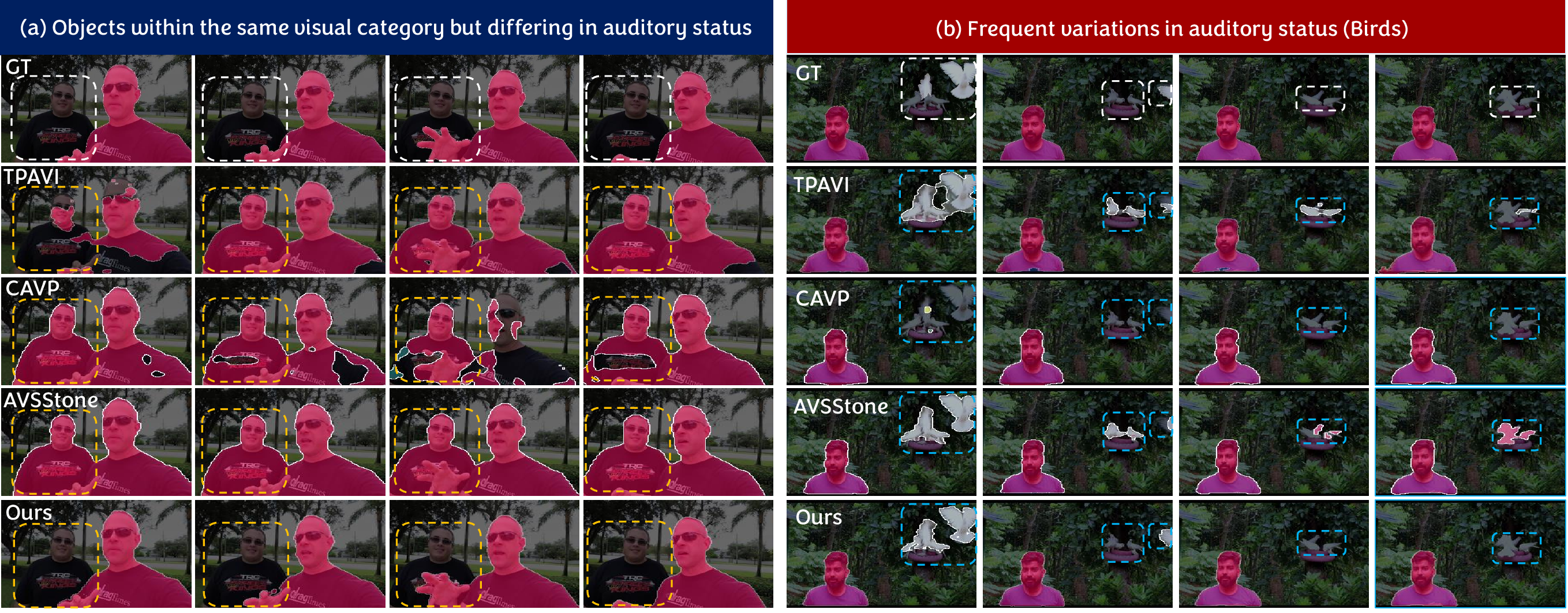}
\end{center}
\vspace{-1.5em}
\caption{Visual comparison of challenging cases (illustrated \ding{182} and \ding{183} in \S \ref{sec:intro}) in AVSBench-Semantic. Refer to \S \ref{sec:qualitative_com} for detailed analysis.}
\label{fig:com_avss}
\vspace{-0.5em}
\end{figure*}

\begin{table*}[htp]
\caption{Quantitative comparisons on $\text{VPO datasets}$ (\S \ref{sec:quantitative_com}). `CPM*' indicates that training and testing are performed on images with the original resolution, while other methods employ images resized to 224 × 224 pixels. The best results are highlighted in \textbf{bold}, and the second-best results are in \underline{underlined}. \colorbox{gray_lc3}{\textcolor{gray_lc2}{Results}} are obtained using our complete design but with different audio-visual backbones.}
\vspace{-0.5em}
\label{tab:vpo}
\centering
\footnotesize
\setlength{\tabcolsep}{1.0mm}
\renewcommand\arraystretch{0.9}
\begin{tabular}{r|c|c|c|ccc|ccc|ccc}
\toprule
\multicolumn{1}{c|}{\multirow{2}{*}{Methods}} & \multicolumn{1}{c|}{\multirow{2}{*}{\shortstack{Visual \\ Backbone}}} & \multicolumn{1}{c|}{\multirow{2}{*}{\shortstack{Audio \\ Backbone}}} & \multicolumn{1}{c|}{\multirow{2}{*}{\shortstack{Trainable \\ Params}}} & \multicolumn{3}{c|}{\textit{VPO-SS}}                                                 & \multicolumn{3}{c|}{\textit{VPO-MS}}                                                 & \multicolumn{3}{c}{\textit{VPO-MSMI}}                                                             \\ \cmidrule{5-13} 
\multicolumn{1}{c|}{}                        & \multicolumn{1}{c|}{}                                 & \multicolumn{1}{c|}{}                                & \multicolumn{1}{c|}{}                                   & \multicolumn{1}{c}{$\mathcal{J}\&\mathcal{F}_\beta \uparrow$} & \multicolumn{1}{c}{$\mathcal{J} \uparrow$} & \multicolumn{1}{c|}{$\mathcal{F}_\beta \uparrow$} & \multicolumn{1}{c}{$\mathcal{J}\&\mathcal{F}_\beta \uparrow$} & \multicolumn{1}{c}{$\mathcal{J} \uparrow$} & \multicolumn{1}{c|}{$\mathcal{F}_\beta \uparrow$} & \multicolumn{1}{c}{$\mathcal{J}\&\mathcal{F}_\beta \uparrow$} & \multicolumn{1}{c}{$\mathcal{J} \uparrow$} & \multicolumn{1}{c}{$\mathcal{F}_\beta \uparrow$}             \\ \midrule

TPAVI  \cite{zhou2022audio}~\textcolor{gray}{\tiny{[ECCV22]}}                                       & PVT-V2-B5                                             & VGGish                                               & 101.32M                                                 & 44.63                     & 41.64                  & 47.62                  & 45.68                     & 42.3                   & 49.06                  & 43.19                     & 40.03                  & \multicolumn{1}{r}{46.34}          \\
AVSegFormer \cite{gao2024avsegformer}~\textcolor{gray}{\tiny{[AAAI24]}}                                 & PVT-V2-B5                                             & VGGish                                               & 186.05M                                                  & 45.94                     & 43.81                  & 48.06                  & 43.72                     & 47.3                   & 40.14                  & 49.93                     & 47.19                  & \multicolumn{1}{r}{52.67}          \\
CAVP \cite{chen2024unraveling}~\textcolor{gray}{\tiny{[CVPR24]}}                                        & ResNet50                                            & VGGish                                               & 119.79M                                                & 67.02                     & 58.81                  & 75.23                  & 61.32                     & 53.24                  & 69.39                  & 56.48                     & 48.18                  & \multicolumn{1}{r}{64.78}          \\
BiasAVS  \cite{sun2024unveiling}~\textcolor{gray}{\tiny{[ECCV24]}}                                   & SwinBase                                              & VGGish                                               & 107.12M                                                 & 67.46                     & 59.14                  & 75.78                  & 63.42                     & 55.61                  & 71.23                  & 57.94                     & 49.6                   & \multicolumn{1}{r}{66.27}          \\
AVSStone  \cite{ma2024stepping} ~\textcolor{gray}{\tiny{[ACM-MM24]}}                                    & SwinBase                                              & VGGish                                               & 114.63M                                                 & \underline{68.54}                     & \underline{59.72}                  & \underline{77.35}                  & \underline{64.26}                     & \underline{56.23}                  & \underline{72.29}                  & \underline{58.76}                     & \underline{50.11}                  & \multicolumn{1}{r}{\underline{67.40}}           \\ \cdashline{1-13}
CPM* \cite{chen2024cpm}~\textcolor{gray}{\tiny{[ECCV24]}}                                         & ResNet50                                              & VGGish                                               & \_                                                      & 73.49                     & 67.09                  & 79.88                  & 72.91                     & 65.91                  & 79.9                   & 68.07                     & 60.55                  & \multicolumn{1}{r}{75.58}          \\ \midrule
\rowcolor[HTML]{f2f2f2} 
{\textcolor{gray_lc2}{\textbf{\textsc{RAVS (Ours)}}}}                                          & \textcolor{gray_lc2}{MIT-B5}                                                 & \textcolor{gray_lc2}{VGGish}                                               & \textcolor{gray_lc2}{103.84M}                                                 & \textcolor{gray_lc2}{74.27}                     & \textcolor{gray_lc2}{67.51}                  & \textcolor{gray_lc2}{81.02}                  & \textcolor{gray_lc2}{72.92}                     & \textcolor{gray_lc2}{66.33}                  & \textcolor{gray_lc2}{79.51}                  & \textcolor{gray_lc2}{68.69}                      & \textcolor{gray_lc2}{61.32}                  & \multicolumn{1}{r}{\textcolor{gray_lc2}{76.05}}          \\
\rowcolor[HTML]{d9d9d9} 
{\textbf{\textsc{RAVS (Ours)}}}                                          & MIT-B5                                                 & HTSAT                                                & 103.84M                                                 & \textbf{74.97}            & \textbf{68.03}         & \textbf{81.90}          & \textbf{73.49}             & \textbf{66.97}         & \textbf{80.01}         & \textbf{69.30}            & \textbf{61.89}         & \multicolumn{1}{r}{\textbf{76.70}} \\ \hline
\end{tabular}
\vspace{-1.5em}
\end{table*}

\subsection{Quantitative Performance}
\label{sec:quantitative_com}
\noindent\textbf{Performance on AVS-Objects and AVS-Semantic.} Table \ref{tab:avss} presents a comparative analysis of our approach against 16 recent audio-visual segmentation (AVS) methods evaluated on AVS-Objects and AVS-Semantic.
In addition to the commonly adopted PVT-V2-B5 \cite{wang2022pvt} and VGGish \cite{chen2020vggsound}, we leverage advanced visual and audio models, MIT-B5 \cite{xie2021segformer} and HTSAT \cite{chen2022hts}, to further enhance segmentation performance. 
Our approach achieves impressive results, with $\mathcal{J} \& \mathcal{F}_\beta$ scores of \textbf{93.0\%} on S4 and \textbf{76.4\%} on MS3 for AVS-Objects, demonstrating precise localization of audible objects.
In particular, on the AVS-Semantic dataset, our method achieves a substantial \textbf{4.2\%} improvement in $\mathcal{J} \& \mathcal{F}_\beta$ over the previous state-of-the-art, AVSStone.
Notably, even with the standard backbones PVT-V2-B5 and VGGish, our model sets new benchmarks across all metrics. 

\noindent\textbf{Performance on VPO.} In Table \ref{tab:vpo}, we report the performance comparison on VPO series datasets.
Our approach greatly outperforms the second-best one AVSStone. 
Specifically, for the VPO-SS dataset, we achieve $\mathcal{J} \& \mathcal{F}_\beta$ scores of \textbf{74.97\%} compared to 68.54\%, for VPO-MS: \textbf{73.48\%} \emph{vs.} 64.26\%, and for VPO-MSMI: \textbf{69.30\%} \emph{vs.} 58.76\%.
Notably, VPO-MSMI includes many cases where objects from the same visual category appear in a frame but have different auditory states \cite{chen2024unraveling}. 
This substantial improvement over AVSStone—$\mathcal{J}$ on VPO-MSMI: \textbf{61.89\%} \emph{vs.} 50.11\% and $\mathcal{F}$: \textbf{76.70\%} \emph{vs.} 67.40\%—demonstrates that our method effectively directs attention to the genuine sounding regions.

\vspace{-0.5em}
\subsection{Qualitative Performance}
\label{sec:qualitative_com}
Fig. \ref{fig:com_avss} presents qualitative comparisons between our method and three competing approaches: TPAVI \cite{zhou2024audio}, CAVP \cite{chen2024unraveling}, and AVSStone \cite{ma2024stepping}. 
In scenario (a), two men are positioned closely but are in different sounding states. 
AVSStone, CAVP, and TPAVI all tend to over-segment the silent man’s region while under-segmenting the audible man’s, reflecting their difficulty in distinguishing between visually similar objects with differing sound states. 
In contrast, our method accurately localizes the speaking man, effectively avoiding these mis-segmentations.
In the more complex scenario (b), the sound status of the bird switches between the second and third frames. 
TPAVI and AVSStone tend to assume the bird is sounding in all frames, leading to segmentation of the bird throughout, while CAVP assumes silence and barely segments the bird region. 
By contrast, our method robustly handles these rapid sound transitions, mitigating these challenges effectively. 


\subsection{Further Analysis}
\label{sec:further_analysis}
\noindent\textbf{Ablation Study.} 
We carry out ablation studies on AVS-Object-MS3 and AVS-Semantic to evaluate the effectiveness of individual components and design choices in our framework. 
Our baseline model includes audio and visual encoders, cross-attention layers for fusion, and a mask decoder. Based on this baseline, we set up two configurations to examine the effectiveness of SGSM.
The first configuration, shown in Row 2 (Baseline+GTF), utilizes ground truth masks to filter silent regions during modality alignment. 
The second configuration, shown in Row 3 (Baseline+SGSM), incorporates our proposed sound-guided semantic merger (SGSM). 
As results indicate, SGSM improves the baseline by 7.4\% in $\mathcal{J} \& \mathcal{F}_\beta$ on AVS-Semantic and achieves 3\% better performance than Baseline+GTF. 
This demonstrates that SGSM effectively directs the model’s attention to audible regions, mitigating interference from silent areas.

Building on SGSM, we introduce contrastive modality alignment (CST), shown in Row 4. 
Compared to SGSM alone, this alignment design improves $\mathcal{J} \& \mathcal{F}_\beta$ by 1.6\% on AVS-Semantic, indicating that aligning compact features with audio representations enhances audio-visual alignment. Additionally, we integrate our uncertainty estimation module (UE) on top of SGSM+CMA, yielding further performance gains. 
On AVS-Semantic, $\mathcal{J}$ and $\mathcal{F}_\beta$ increase by 1.7\% and 1.3\%, respectively, confirming that UE effectively mitigates errors caused by frequent changes in an object’s audible state, resulting in more accurate segmentation.

\begin{figure}[t]
\begin{center}
\includegraphics[width=1.0\linewidth]{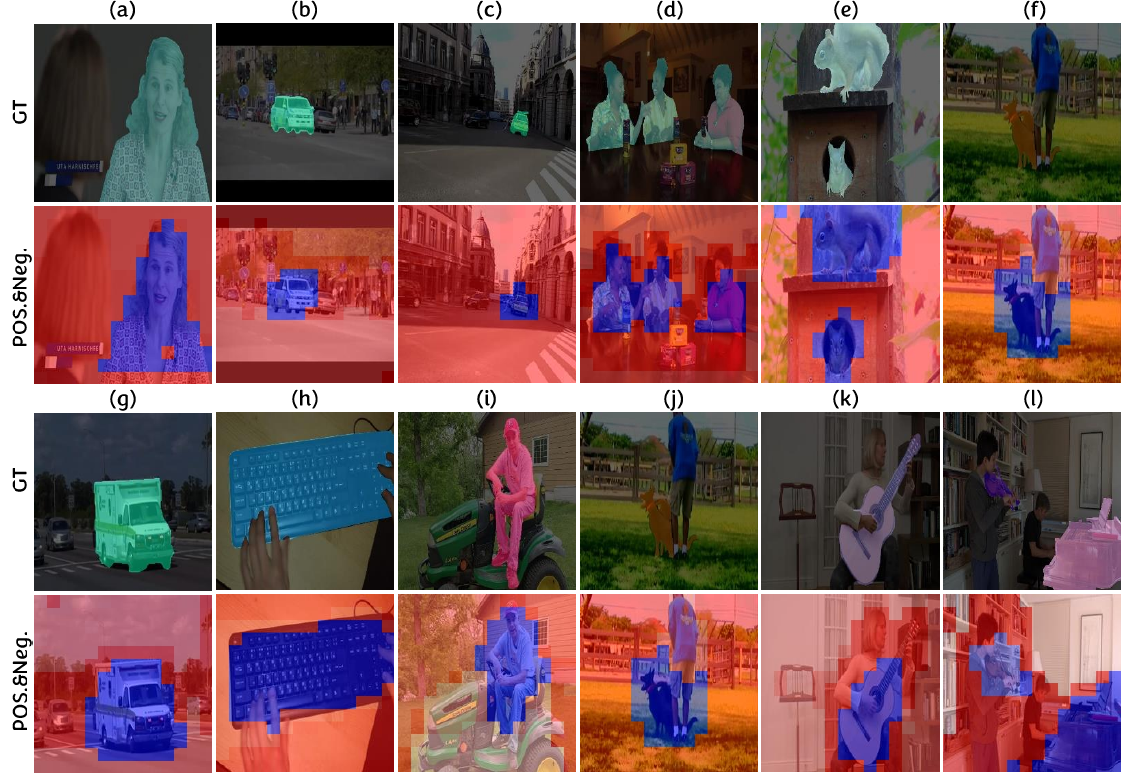}
\end{center}
\vspace{-1.5em}
\caption{Visualization of positive ({Pos.}) and negative ({Neg.}) samples generated by the AMA module. \textcolor{red}{Red} indicates negative samples, while \textcolor{blue}{blue} represents positive samples. Darker blue indicates regions with a higher responsiveness to audio cues, while darker red indicates lower responsiveness. See detailed analysis in \S\ref{sec:further_analysis}.}
\label{fig:cst_vis}
\vspace{-1.0em}
\end{figure}

\noindent\textbf{Effectiveness of Audio Guided Alignment.}
In Fig. \ref{fig:cst_vis}, we map the compact representation identified as a positive or negative sample back onto the image based on the locations of their visual combinations and show their responsiveness to the audio cues. 
When two objects of the same visual category but different sound states appear in a video frame, our method effectively identifies the truly sounding object as a positive sample and the silent object as a negative sample. 
For example, in (a)-(c), in scenes with two women where only one is speaking, and in scenes with multiple adjacent cars where only one is making sounds, our method accurately associates the sound with the truly audible objects.

When multiple objects of the same category are sounding simultaneously, our method consistently classifies all sounding objects as positive samples, as shown in (d)-(f).
Additionally, by analyzing the sound responsiveness of all samples, we observe that regions around positive samples typically show lower sound responsiveness.  
This indicates that the audio-guided feature merging not only focuses effectively on relevant sounding regions but also assists in identifying hard samples around the sounding regions.

\begin{table}[t]
\caption{Evaluation of component effectiveness (\S \ref{sec:further_analysis}). GTF: Ground-Truth mask filtering; SGSM: sound-guided semantic merging; CST: contrastive-based alignment; AMA: audio-guided modality alignment; UE: uncertainty estimation.}
\vspace{-0.5em}
\label{tab:ablation}
\centering
\footnotesize
\setlength{\tabcolsep}{0.7mm}
\renewcommand\arraystretch{1.0}
\begin{tabular}{l|ccc|ccc}
\midrule
\multirow{2}{*}{Settings}    & \multicolumn{3}{c|}{\textit{AVS-Objects-MS3}                                     } & \multicolumn{3}{c}{\textit{AVS-Semantic}}                                         \\ \cmidrule{2-7} 
                           & \multicolumn{1}{c}{$\mathcal{J}\&\mathcal{F}_\beta \uparrow$} & \multicolumn{1}{c}{$\mathcal{J} \uparrow$} & \multicolumn{1}{c|}{$\mathcal{F}_\beta \uparrow$} & \multicolumn{1}{c}{$\mathcal{J}\&\mathcal{F}_\beta \uparrow$} & \multicolumn{1}{c}{$\mathcal{J} \uparrow$} & \multicolumn{1}{c}{$\mathcal{F}_\beta \uparrow$}\\ \hline
Baseline                  & 64.3          & 59.8     & 68.7     & 57.2     & 51.4 & 62.9                     \\ \hline
+GTF               & 67.9          & 62.2     & 73.6     & 59.6     & 55.1 & 64.1                     \\ \hline
+SGSM             & 70.9          & 65.3     & 76.4     & 62.6     & 57.7 & 67.5                     \\
+SGSM+CST (AMA)    & 72.2          & 66.1     & 78.3     & 64.2     & 59.1 & 69.3                     \\ \hline
\rowcolor[HTML]{d9d9d9} 
+AMA+UE (Ours)          & \textbf{76.4}                     & \textbf{70.6}                  & \textbf{82.1}                   & \textbf{65.7}                     & \textbf{60.8}                 & \textbf{70.6}                    \\ \bottomrule
\end{tabular}
\vspace{-2em}
\end{table}

\begin{figure}[]
\begin{center}
\includegraphics[width=1.0\linewidth]{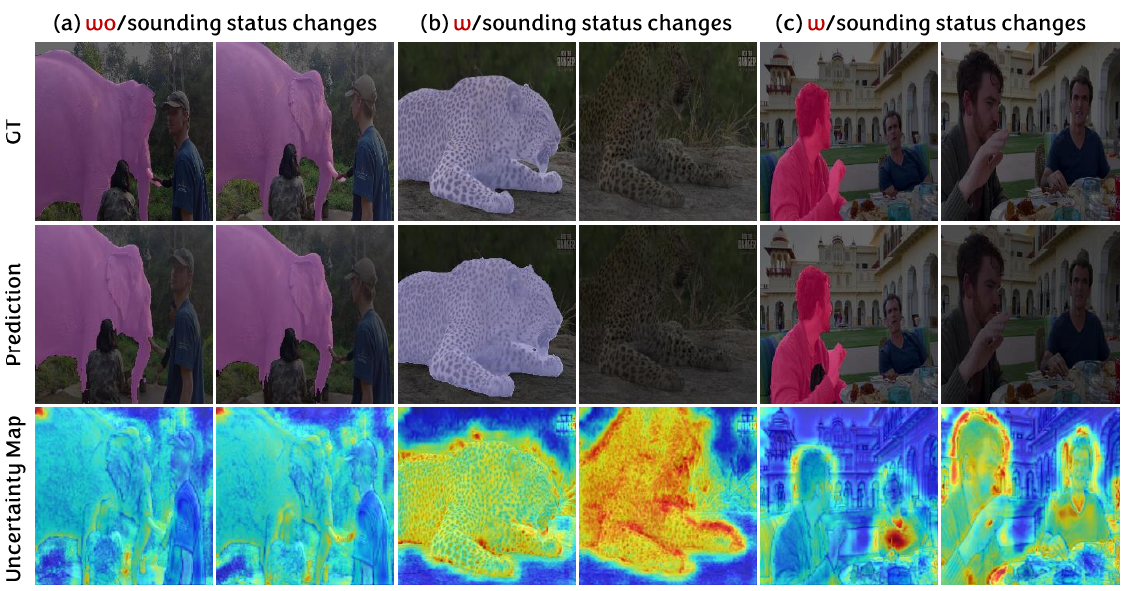}
\end{center}
\vspace{-1.5em}
\caption{Visualizations of uncertainty maps, with colors ranging from red (high uncertainty) to blue (low uncertainty). }
\label{fig:uncertainty_vis}
\vspace{-1.0em}
\end{figure}

\noindent\textbf{Effectiveness of Uncertainty Estimation.}
Fig. \ref{fig:uncertainty_vis} presents the visualizations of uncertainty maps generated by our uncertainty estimation (UE) module.
In (a), where the sound state of all objects remains constant, the model shows consistently low uncertainty values across the entire image, indicating high confidence in its segmentation predictions. 
In contrast, in (b), where a \texttt{`leopard'} undergoes a transition in the sound state between adjacent frames, the model exhibits higher uncertainty for both sound-emitting and silent frames, with even higher uncertainty values in the silent frames. 
This higher uncertainty effectively lowers the prediction probability in the \texttt{`leopard'} region in the second frame, reducing over-segmentation in silent frames.
These findings demonstrate that frequent sound state transitions significantly impact the model's confidence in its predictions, suggesting that uncertainty estimation is an effective way to mitigate mis-segmentation issues.

\noindent\textbf{Effects of Group Number.}
We conduct a study on the impact of different group numbers when conducting audio-guided alignment.
The grouping setting of [14, 7, 5]  (approximately 1/4 of the feature map size) achieves the best performance, attaining the highest $\mathcal{J}\&\mathcal{F}_\beta$ scores on both datasets. 
In contrast, smaller grouping numbers like [5, 5, 5] led to decreased performance.  
This suggests that progressively reducing the group number supports more effective semantic merging of sound-emitting regions under audio guidance. 
Additionally, a larger group number, such as [28, 14, 5], divides the feature map into many small groups, dispersing attention across numerous small regions and thus leading to poorer results.

\begin{table}[]
\caption{Comparison of different group numbers for three-scale feature maps. See analysis in \S\ref{sec:further_analysis}.}
\vspace{-1em}
\label{tab:cluster_ana}
\centering
\footnotesize
\setlength{\tabcolsep}{1.4mm}
\renewcommand\arraystretch{0.9}
\begin{tabular}{l|ccc|ccc}
\midrule
\multirow{2}{*}{Groups}    & \multicolumn{3}{c|}{\textit{AVS-Objects-MS3}                                     } & \multicolumn{3}{c}{\textit{AVS-Semantic}}                                         \\ \cmidrule{2-7} 
                           & \multicolumn{1}{c}{$\mathcal{J}\&\mathcal{F}_\beta \uparrow$} & \multicolumn{1}{c}{$\mathcal{J} \uparrow$} & \multicolumn{1}{c|}{$\mathcal{F}_\beta \uparrow$} & \multicolumn{1}{c}{$\mathcal{J}\&\mathcal{F}_\beta \uparrow$} & \multicolumn{1}{c}{$\mathcal{J} \uparrow$} & \multicolumn{1}{c}{$\mathcal{F}_\beta \uparrow$}\\ \midrule
{[}28, 14, 5{]}  & 73.7                     & 68.6                  & 78.8                   & 63.4                     & 57.8                  & 68.9                  \\
\rowcolor[HTML]{d9d9d9} 
{[}14, 7, 5{]}             & \textbf{76.4}                     & \textbf{70.6}                  & \textbf{82.1}                   & \textbf{65.7}                     & \textbf{60.8}                 & \textbf{70.6}                  \\

{[}10, 7, 5{]}            & 75.1                    & 69.3                 & 80.8                    & 64.4                   & 58.8        & 70.0         \\
{[}5,5,5{]}                & 71.5                     & 66.7                  & 76.3                   & 62.6                     & 56.5                  & 68.7                  \\ \bottomrule
\end{tabular}
\vspace{-2em}
\end{table}

\section{Conclusion}
\label{Sec:conclusion}
In this paper, we propose a novel framework to address modality misalignment caused by ambiguous spatio-temporal correspondences. 
Unlike previous methods that rely on global attention or employ masks to exclude silent regions, our approach first utilizes audio to guide the merging of sounding regions, followed by contrastive learning to align compact visual representations with audio cues.
This design enhances the model's ability to distinguish between audible and silent objects, even when visually similar but acoustically different objects are in close proximity. 
Additionally, we introduce uncertainty estimation to mitigate predictive instability arising from frequent changes in an object's sound state. 
The two items enable our framework to prevent the over-segmentation of silent regions in complex scenarios. 
Extensive experiments demonstrate the effectiveness of our approach in handling challenging cases.

\vspace{0.5em}
\noindent\textbf{Acknowledgements.} This work is supported by ARC-Discovery (DP220100800 to XY), and ARC-DECRA  (DE230100477 to XY). Chen Liu is funded by the China Scholarship Council and CSIRO top-up (50092128). 

\clearpage
{
    \small
    \bibliographystyle{ieeenat_fullname}
    \bibliography{main}

\begin{thebibliography}{50}
\providecommand{\natexlab}[1]{#1}
\providecommand{\url}[1]{\texttt{#1}}
\expandafter\ifx\csname urlstyle\endcsname\relax
  \providecommand{\doi}[1]{doi: #1}\else
  \providecommand{\doi}{doi: \begingroup \urlstyle{rm}\Url}\fi

\bibitem[Chen et~al.(2020{\natexlab{a}})Chen, Xie, Vedaldi, and Zisserman]{chen2020vggsound}
Honglie Chen, Weidi Xie, Andrea Vedaldi, and Andrew Zisserman.
\newblock Vggsound: A large-scale audio-visual dataset.
\newblock In \emph{ICASSP 2020-2020 IEEE International Conference on Acoustics, Speech and Signal Processing (ICASSP)}, pages 721--725. IEEE, 2020{\natexlab{a}}.

\bibitem[Chen et~al.(2021)Chen, Xie, Afouras, Nagrani, Vedaldi, and Zisserman]{chen2021localizing}
Honglie Chen, Weidi Xie, Triantafyllos Afouras, Arsha Nagrani, Andrea Vedaldi, and Andrew Zisserman.
\newblock Localizing visual sounds the hard way.
\newblock In \emph{Proceedings of the IEEE/CVF conference on computer vision and pattern recognition}, pages 16867--16876, 2021.

\bibitem[Chen et~al.(2022)Chen, Du, Zhu, Ma, Berg-Kirkpatrick, and Dubnov]{chen2022hts}
Ke Chen, Xingjian Du, Bilei Zhu, Zejun Ma, Taylor Berg-Kirkpatrick, and Shlomo Dubnov.
\newblock Hts-at: A hierarchical token-semantic audio transformer for sound classification and detection.
\newblock In \emph{ICASSP 2022-2022 IEEE International Conference on Acoustics, Speech and Signal Processing (ICASSP)}, pages 646--650. IEEE, 2022.

\bibitem[Chen et~al.(2020{\natexlab{b}})Chen, Kornblith, Norouzi, and Hinton]{chen2020simple}
Ting Chen, Simon Kornblith, Mohammad Norouzi, and Geoffrey Hinton.
\newblock A simple framework for contrastive learning of visual representations.
\newblock In \emph{International conference on machine learning}, pages 1597--1607. PMLR, 2020{\natexlab{b}}.

\bibitem[Chen et~al.(2024{\natexlab{a}})Chen, Liu, Wang, Liu, Wang, Frazer, and Carneiro]{chen2024unraveling}
Yuanhong Chen, Yuyuan Liu, Hu Wang, Fengbei Liu, Chong Wang, Helen Frazer, and Gustavo Carneiro.
\newblock Unraveling instance associations: A closer look for audio-visual segmentation.
\newblock In \emph{Proceedings of the IEEE/CVF Conference on Computer Vision and Pattern Recognition}, pages 26497--26507, 2024{\natexlab{a}}.

\bibitem[Chen et~al.(2024{\natexlab{b}})Chen, Wang, Liu, Wang, and Carneiro]{chen2024cpm}
Yuanhong Chen, Chong Wang, Yuyuan Liu, Hu Wang, and Gustavo Carneiro.
\newblock Cpm: Class-conditional prompting machine for audio-visual segmentation.
\newblock \emph{arXiv preprint arXiv:2407.05358}, 2024{\natexlab{b}}.

\bibitem[Cheng et~al.(2021)Cheng, Choudhuri, Misra, Kirillov, Girdhar, and Schwing]{cheng2021mask2formervideo}
Bowen Cheng, Anwesa Choudhuri, Ishan Misra, Alexander Kirillov, Rohit Girdhar, and Alexander~G. Schwing.
\newblock Mask2former for video instance segmentation, 2021.

\bibitem[Du et~al.(2016)Du, Ding, and Jia]{du2016study}
Mingjing Du, Shifei Ding, and Hongjie Jia.
\newblock Study on density peaks clustering based on k-nearest neighbors and principal component analysis.
\newblock \emph{Knowledge-Based Systems}, 99:\penalty0 135--145, 2016.

\bibitem[Gao et~al.(2024)Gao, Chen, Chen, Wang, and Lu]{gao2024avsegformer}
Shengyi Gao, Zhe Chen, Guo Chen, Wenhai Wang, and Tong Lu.
\newblock Avsegformer: Audio-visual segmentation with transformer.
\newblock In \emph{Proceedings of the AAAI Conference on Artificial Intelligence}, pages 12155--12163, 2024.

\bibitem[Grill et~al.(2020)Grill, Strub, Altch{\'e}, Tallec, Richemond, Buchatskaya, Doersch, Avila~Pires, Guo, Gheshlaghi~Azar, et~al.]{grill2020bootstrap}
Jean-Bastien Grill, Florian Strub, Florent Altch{\'e}, Corentin Tallec, Pierre Richemond, Elena Buchatskaya, Carl Doersch, Bernardo Avila~Pires, Zhaohan Guo, Mohammad Gheshlaghi~Azar, et~al.
\newblock Bootstrap your own latent-a new approach to self-supervised learning.
\newblock \emph{Advances in neural information processing systems}, 33:\penalty0 21271--21284, 2020.

\bibitem[Hao et~al.(2024)Hao, Mao, He, Han, Dai, and Zhong]{hao2024improving}
Dawei Hao, Yuxin Mao, Bowen He, Xiaodong Han, Yuchao Dai, and Yiran Zhong.
\newblock Improving audio-visual segmentation with bidirectional generation.
\newblock In \emph{Proceedings of the AAAI Conference on Artificial Intelligence}, pages 2067--2075, 2024.

\bibitem[Hu et~al.(2020)Hu, Qian, Jiang, Tan, Wen, Ding, Lin, and Dou]{hu2020discriminative}
Di Hu, Rui Qian, Minyue Jiang, Xiao Tan, Shilei Wen, Errui Ding, Weiyao Lin, and Dejing Dou.
\newblock Discriminative sounding objects localization via self-supervised audiovisual matching.
\newblock \emph{Advances in Neural Information Processing Systems}, 33:\penalty0 10077--10087, 2020.

\bibitem[Huang et~al.(2023)Huang, Li, Wang, Zhu, Dai, Han, Rong, and Liu]{huang2023discovering}
Shaofei Huang, Han Li, Yuqing Wang, Hongji Zhu, Jiao Dai, Jizhong Han, Wenge Rong, and Si Liu.
\newblock Discovering sounding objects by audio queries for audio visual segmentation.
\newblock \emph{arXiv preprint arXiv:2309.09501}, 2023.

\bibitem[King et~al.(2021)King, Tatoglu, Iglesias, and Matriss]{king2021audio}
EA King, A Tatoglu, D Iglesias, and A Matriss.
\newblock Audio-visual based non-line-of-sight sound source localization: A feasibility study.
\newblock \emph{Applied Acoustics}, 171:\penalty0 107674, 2021.

\bibitem[Le-Khac et~al.(2020)Le-Khac, Healy, and Smeaton]{le2020contrastive}
Phuc~H Le-Khac, Graham Healy, and Alan~F Smeaton.
\newblock Contrastive representation learning: A framework and review.
\newblock \emph{Ieee Access}, 8:\penalty0 193907--193934, 2020.

\bibitem[Li et~al.(2023)Li, Yang, Chen, Yang, and Xiao]{li2023catr}
Kexin Li, Zongxin Yang, Lei Chen, Yi Yang, and Jun Xiao.
\newblock Catr: Combinatorial-dependence audio-queried transformer for audio-visual video segmentation.
\newblock In \emph{Proceedings of the 31st ACM International Conference on Multimedia}, pages 1485--1494, 2023.

\bibitem[Li et~al.(2019)Li, Sun, Meng, Liang, Wu, and Li]{li2019dice}
Xiaoya Li, Xiaofei Sun, Yuxian Meng, Junjun Liang, Fei Wu, and Jiwei Li.
\newblock Dice loss for data-imbalanced nlp tasks.
\newblock \emph{arXiv preprint arXiv:1911.02855}, 2019.

\bibitem[Li et~al.(2024)Li, Wang, Xu, Peng, Singh, Lu, and Raj]{li2024qdformer}
Xiang Li, Jinglu Wang, Xiaohao Xu, Xiulian Peng, Rita Singh, Yan Lu, and Bhiksha Raj.
\newblock Qdformer: Towards robust audiovisual segmentation in complex environments with quantization-based semantic decomposition.
\newblock In \emph{Proceedings of the IEEE/CVF Conference on Computer Vision and Pattern Recognition}, pages 3402--3413, 2024.

\bibitem[Lin et~al.(2014)Lin, Maire, Belongie, Hays, Perona, Ramanan, Doll{\'a}r, and Zitnick]{lin2014microsoft}
Tsung-Yi Lin, Michael Maire, Serge Belongie, James Hays, Pietro Perona, Deva Ramanan, Piotr Doll{\'a}r, and C~Lawrence Zitnick.
\newblock Microsoft coco: Common objects in context.
\newblock In \emph{Computer Vision--ECCV 2014: 13th European Conference, Zurich, Switzerland, September 6-12, 2014, Proceedings, Part V 13}, pages 740--755. Springer, 2014.

\bibitem[Liu et~al.(2023{\natexlab{a}})Liu, Li, Qi, Zhang, Li, Wang, and Yu]{liu2023audio}
Chen Liu, Peike~Patrick Li, Xingqun Qi, Hu Zhang, Lincheng Li, Dadong Wang, and Xin Yu.
\newblock Audio-visual segmentation by exploring cross-modal mutual semantics.
\newblock In \emph{Proceedings of the 31st ACM International Conference on Multimedia}, pages 7590--7598, 2023{\natexlab{a}}.

\bibitem[Liu et~al.(2024{\natexlab{a}})Liu, Li, Zhang, Li, Huang, Wang, and Yu]{liu2024bavs}
Chen Liu, Peike Li, Hu Zhang, Lincheng Li, Zi Huang, Dadong Wang, and Xin Yu.
\newblock Bavs: bootstrapping audio-visual segmentation by integrating foundation knowledge.
\newblock \emph{IEEE Transactions on Multimedia}, 2024{\natexlab{a}}.

\bibitem[Liu et~al.(2023{\natexlab{b}})Liu, Ju, Ma, Wang, Wang, and Zhang]{liu2023audio_a}
Jinxiang Liu, Chen Ju, Chaofan Ma, Yanfeng Wang, Yu Wang, and Ya Zhang.
\newblock Audio-aware query-enhanced transformer for audio-visual segmentation.
\newblock \emph{arXiv preprint arXiv:2307.13236}, 2023{\natexlab{b}}.

\bibitem[Liu et~al.(2024{\natexlab{b}})Liu, Wang, Ju, Ma, Zhang, and Xie]{liu2024annotation}
Jinxiang Liu, Yu Wang, Chen Ju, Chaofan Ma, Ya Zhang, and Weidi Xie.
\newblock Annotation-free audio-visual segmentation.
\newblock In \emph{Proceedings of the IEEE/CVF Winter Conference on Applications of Computer Vision}, pages 5604--5614, 2024{\natexlab{b}}.

\bibitem[Ma et~al.(2024)Ma, Sun, Wang, and Hu]{ma2024stepping}
Juncheng Ma, Peiwen Sun, Yaoting Wang, and Di Hu.
\newblock Stepping stones: A progressive training strategy for audio-visual semantic segmentation.
\newblock \emph{arXiv preprint arXiv:2407.11820}, 2024.

\bibitem[Mahmud et~al.(2024)Mahmud, Tian, and Marculescu]{mahmud2024t}
Tanvir Mahmud, Yapeng Tian, and Diana Marculescu.
\newblock T-vsl: Text-guided visual sound source localization in mixtures.
\newblock In \emph{Proceedings of the IEEE/CVF Conference on Computer Vision and Pattern Recognition}, pages 26742--26751, 2024.

\bibitem[Mao et~al.(2023{\natexlab{a}})Mao, Zhang, Xiang, Lv, Zhong, and Dai]{mao2023contrastive}
Yuxin Mao, Jing Zhang, Mochu Xiang, Yunqiu Lv, Yiran Zhong, and Yuchao Dai.
\newblock Contrastive conditional latent diffusion for audio-visual segmentation.
\newblock \emph{arXiv preprint arXiv:2307.16579}, 2023{\natexlab{a}}.

\bibitem[Mao et~al.(2023{\natexlab{b}})Mao, Zhang, Xiang, Zhong, and Dai]{mao2023multimodal}
Yuxin Mao, Jing Zhang, Mochu Xiang, Yiran Zhong, and Yuchao Dai.
\newblock Multimodal variational auto-encoder based audio-visual segmentation.
\newblock In \emph{Proceedings of the IEEE/CVF International Conference on Computer Vision}, pages 954--965, 2023{\natexlab{b}}.

\bibitem[Masuyama et~al.(2020)Masuyama, Bando, Yatabe, Sasaki, Onishi, and Oikawa]{masuyama2020self}
Yoshiki Masuyama, Yoshiaki Bando, Kohei Yatabe, Yoko Sasaki, Masaki Onishi, and Yasuhiro Oikawa.
\newblock Self-supervised neural audio-visual sound source localization via probabilistic spatial modeling.
\newblock In \emph{2020 IEEE/RSJ International Conference on Intelligent Robots and Systems (IROS)}, pages 4848--4854. IEEE, 2020.

\bibitem[Minka(2000)]{minka2000estimating}
Thomas Minka.
\newblock Estimating a dirichlet distribution, 2000.

\bibitem[Mo and Morgado(2022{\natexlab{a}})]{mo2022closer}
Shentong Mo and Pedro Morgado.
\newblock A closer look at weakly-supervised audio-visual source localization.
\newblock \emph{Advances in Neural Information Processing Systems}, 35:\penalty0 37524--37536, 2022{\natexlab{a}}.

\bibitem[Mo and Morgado(2022{\natexlab{b}})]{mo2022localizing}
Shentong Mo and Pedro Morgado.
\newblock Localizing visual sounds the easy way.
\newblock In \emph{European Conference on Computer Vision}, pages 218--234. Springer, 2022{\natexlab{b}}.

\bibitem[Mo and Tian(2023)]{mo2023audio}
Shentong Mo and Yapeng Tian.
\newblock Audio-visual grouping network for sound localization from mixtures.
\newblock In \emph{Proceedings of the IEEE/CVF Conference on Computer Vision and Pattern Recognition}, pages 10565--10574, 2023.

\bibitem[Oord et~al.(2018)Oord, Li, and Vinyals]{oord2018representation}
Aaron van~den Oord, Yazhe Li, and Oriol Vinyals.
\newblock Representation learning with contrastive predictive coding.
\newblock \emph{arXiv preprint arXiv:1807.03748}, 2018.

\bibitem[Qian et~al.(2020)Qian, Hu, Dinkel, Wu, Xu, and Lin]{qian2020multiple}
Rui Qian, Di Hu, Heinrich Dinkel, Mengyue Wu, Ning Xu, and Weiyao Lin.
\newblock Multiple sound sources localization from coarse to fine.
\newblock In \emph{Computer Vision--ECCV 2020: 16th European Conference, Glasgow, UK, August 23--28, 2020, Proceedings, Part XX 16}, pages 292--308. Springer, 2020.

\bibitem[Rachavarapu et~al.(2021)Rachavarapu, Sundaresha, Rajagopalan, et~al.]{rachavarapu2021localize}
Kranthi~Kumar Rachavarapu, Vignesh Sundaresha, AN Rajagopalan, et~al.
\newblock Localize to binauralize: Audio spatialization from visual sound source localization.
\newblock In \emph{Proceedings of the IEEE/CVF International Conference on Computer Vision}, pages 1930--1939, 2021.

\bibitem[Sensoy et~al.(2018)Sensoy, Kaplan, and Kandemir]{sensoy2018evidential}
Murat Sensoy, Lance Kaplan, and Melih Kandemir.
\newblock Evidential deep learning to quantify classification uncertainty.
\newblock \emph{Advances in neural information processing systems}, 31, 2018.

\bibitem[Sun et~al.(2024)Sun, Zhang, and Hu]{sun2024unveiling}
Peiwen Sun, Honggang Zhang, and Di Hu.
\newblock Unveiling and mitigating bias in audio visual segmentation.
\newblock \emph{arXiv preprint arXiv:2407.16638}, 2024.

\bibitem[Sun et~al.(2023)Sun, Zhang, Wang, Liu, Zhong, Feng, Guo, Zhang, and Barnes]{sun2023learning}
Weixuan Sun, Jiayi Zhang, Jianyuan Wang, Zheyuan Liu, Yiran Zhong, Tianpeng Feng, Yandong Guo, Yanhao Zhang, and Nick Barnes.
\newblock Learning audio-visual source localization via false negative aware contrastive learning.
\newblock In \emph{Proceedings of the IEEE/CVF Conference on Computer Vision and Pattern Recognition}, pages 6420--6429, 2023.

\bibitem[Tian et~al.(2018)Tian, Shi, Li, Duan, and Xu]{tian2018audio}
Yapeng Tian, Jing Shi, Bochen Li, Zhiyao Duan, and Chenliang Xu.
\newblock Audio-visual event localization in unconstrained videos.
\newblock In \emph{Proceedings of the European conference on computer vision (ECCV)}, pages 247--263, 2018.

\bibitem[Wang et~al.(2022)Wang, Xie, Li, Fan, Song, Liang, Lu, Luo, and Shao]{wang2022pvt}
Wenhai Wang, Enze Xie, Xiang Li, Deng-Ping Fan, Kaitao Song, Ding Liang, Tong Lu, Ping Luo, and Ling Shao.
\newblock Pvt v2: Improved baselines with pyramid vision transformer.
\newblock \emph{Computational Visual Media}, 8\penalty0 (3):\penalty0 415--424, 2022.

\bibitem[Wang et~al.(2024)Wang, Liu, Li, Ding, Hu, and Li]{wang2024prompting}
Yaoting Wang, Weisong Liu, Guangyao Li, Jian Ding, Di Hu, and Xi Li.
\newblock Prompting segmentation with sound is generalizable audio-visual source localizer.
\newblock In \emph{Proceedings of the AAAI Conference on Artificial Intelligence}, pages 5669--5677, 2024.

\bibitem[Xia and Zhao(2022)]{xia2022cross}
Yan Xia and Zhou Zhao.
\newblock Cross-modal background suppression for audio-visual event localization.
\newblock In \emph{Proceedings of the IEEE/CVF conference on computer vision and pattern recognition}, pages 19989--19998, 2022.

\bibitem[Xie et~al.(2021)Xie, Wang, Yu, Anandkumar, Alvarez, and Luo]{xie2021segformer}
Enze Xie, Wenhai Wang, Zhiding Yu, Anima Anandkumar, Jose~M Alvarez, and Ping Luo.
\newblock Segformer: Simple and efficient design for semantic segmentation with transformers.
\newblock \emph{Advances in neural information processing systems}, 34:\penalty0 12077--12090, 2021.

\bibitem[Yan et~al.(2024)Yan, Zhang, Guo, Chen, Zhang, Li, Qiao, Dong, He, and Gao]{yan2024referred}
Shilin Yan, Renrui Zhang, Ziyu Guo, Wenchao Chen, Wei Zhang, Hongyang Li, Yu Qiao, Hao Dong, Zhongjiang He, and Peng Gao.
\newblock Referred by multi-modality: A unified temporal transformer for video object segmentation.
\newblock In \emph{Proceedings of the AAAI Conference on Artificial Intelligence}, pages 6449--6457, 2024.

\bibitem[Yang et~al.(2024)Yang, Nie, Li, Gao, Guo, Zhen, Yan, and Xiang]{yang2024cooperation}
Qi Yang, Xing Nie, Tong Li, Pengfei Gao, Ying Guo, Cheng Zhen, Pengfei Yan, and Shiming Xiang.
\newblock Cooperation does matter: Exploring multi-order bilateral relations for audio-visual segmentation.
\newblock In \emph{Proceedings of the IEEE/CVF Conference on Computer Vision and Pattern Recognition}, pages 27134--27143, 2024.

\bibitem[Ying et~al.(2023)Ying, Zhong, Mao, Wang, Chen, Wu, Liu, Fan, Zhuge, and Shen]{ying2023ctvis}
Kaining Ying, Qing Zhong, Weian Mao, Zhenhua Wang, Hao Chen, Lin~Yuanbo Wu, Yifan Liu, Chengxiang Fan, Yunzhi Zhuge, and Chunhua Shen.
\newblock Ctvis: Consistent training for online video instance segmentation.
\newblock In \emph{Proceedings of the IEEE/CVF International Conference on Computer Vision}, pages 899--908, 2023.

\bibitem[Yu et~al.(2016)Yu, Jiang, Wang, Cao, and Huang]{yu2016unitbox}
Jiahui Yu, Yuning Jiang, Zhangyang Wang, Zhimin Cao, and Thomas Huang.
\newblock Unitbox: An advanced object detection network.
\newblock In \emph{Proceedings of the 24th ACM international conference on Multimedia}, pages 516--520, 2016.

\bibitem[Zeng et~al.(2022)Zeng, Jin, Liu, Qian, Luo, Ouyang, and Wang]{zeng2022not}
Wang Zeng, Sheng Jin, Wentao Liu, Chen Qian, Ping Luo, Wanli Ouyang, and Xiaogang Wang.
\newblock Not all tokens are equal: Human-centric visual analysis via token clustering transformer.
\newblock In \emph{Proceedings of the IEEE/CVF conference on computer vision and pattern recognition}, pages 11101--11111, 2022.

\bibitem[Zhou et~al.(2022)Zhou, Wang, Zhang, Sun, Zhang, Birchfield, Guo, Kong, Wang, and Zhong]{zhou2022audio}
Jinxing Zhou, Jianyuan Wang, Jiayi Zhang, Weixuan Sun, Jing Zhang, Stan Birchfield, Dan Guo, Lingpeng Kong, Meng Wang, and Yiran Zhong.
\newblock Audio--visual segmentation.
\newblock In \emph{European Conference on Computer Vision}, pages 386--403. Springer, 2022.

\bibitem[Zhou et~al.(2024)Zhou, Shen, Wang, Zhang, Sun, Zhang, Birchfield, Guo, Kong, Wang, et~al.]{zhou2024audio}
Jinxing Zhou, Xuyang Shen, Jianyuan Wang, Jiayi Zhang, Weixuan Sun, Jing Zhang, Stan Birchfield, Dan Guo, Lingpeng Kong, Meng Wang, et~al.
\newblock Audio-visual segmentation with semantics.
\newblock \emph{International Journal of Computer Vision}, pages 1--21, 2024.

\end{thebibliography}
}
\end{document}